\documentclass[aps, preprintnumbers,amsmath,amssymb,nofootinbib,notitlepage, onecolumn]{revtex4-1}
\pdfoutput=1

\usepackage{amssymb}
\usepackage{graphicx}
\usepackage{amsmath}
\usepackage{multirow}
\usepackage{verbatim}
\usepackage[usenames,dvipsnames]{color}
\newcommand{\nc}{\newcommand}
\nc{\av}[1]{\textcolor{RoyalBlue}{{\bf [AV: #1] }}}
\nc{\lo}[1]{\textcolor{Magenta}{{\bf [LLH: #1] }}}
\nc{\om}[1]{\textcolor{ForestGreen}{{\bf [OM: #1] }}}
\nc{\spr}[1]{\textcolor{Red}{{\bf [SPR: #1] }}}

\nc{\ba}{\begin{eqnarray}}
\nc{\ea}{\end{eqnarray}}
\nc{\bc}{\begin{center}}
\nc{\ec}{\end{center}}
\nc{\ie}{\textit{i.e.}}
\nc{\del}{\partial}
\nc{\ud}{\mathrm{d}}
\nc{\mx}{m_\chi}
\nc{\feff}{f_{\mathrm{eff}}}
\nc{\geff}{g_{\mathrm{eff}}}
\nc{\zeff}{z_{\mathrm{eff}}}
\nc{\zmin}{z_{\mathrm{min}}}
\nc{\zmax}{z_{\mathrm{max}}}
\nc{\sv}{ \sigma v}
\nc{\sva}{\langle \sigma v \rangle}
\nc{\zkd}{z_\mathrm{KD}}
\nc{\Tkd}{T_\mathrm{KD}}
\nc{\Tcd}{T_{\rm FO} }
\nc{\zcd}{z_{\rm FO}}
\nc{\zref}{z_{\rm ref}}
\nc{\svref}{\sv_{\rm ref}}
\nc{\vref}{v_{\rm ref}}
\nc{\vcd}{v_{\rm FO}}

\begin{document}

\preprint{IFIC/13--54}

\title{Constraining dark matter late--time energy injection: \\ decays
  and p--wave annihilations}
\author{Roberta Diamanti$^1$, Laura Lopez-Honorez$^2$, Olga Mena$^1$,
  Sergio Palomares-Ruiz$^{1}$ and Aaron C. Vincent$^1$} 
  \affiliation{$^1$Instituto de F\'{\i}sica Corpuscular (IFIC)$,$
  CSIC-Universitat de Val\`encia$,$ \\  
Apartado de Correos 22085$,$ E-46071 Valencia$,$ Spain}
\affiliation{$^2$
Theoretische Natuurkunde Vrije Universiteit Brussel and The
International Solvay Institutes Pleinlaan 2$,$  B-1050 Brussels$,$
Belgium}

\begin{abstract}
We use the latest cosmic microwave background (CMB) observations to
provide updated constraints on the dark matter lifetime as well as on
p--wave suppressed annihilation cross sections in the 1~MeV to 1~TeV
mass range.  In contrast to scenarios with an s--wave dominated
annihilation cross section, which mainly affect the CMB close to the
last scattering surface, signatures associated with these scenarios
essentially appear at low redshifts ($z \lesssim 50$) when structure   
began to form, and thus manifest at lower multipoles in the CMB power
spectrum.  We use data from Planck, WMAP9, SPT and ACT, as well as
Lyman--$\alpha$ measurements of the matter temperature at $z \sim 4$
to set a 95\% confidence level lower bound on the dark matter lifetime
of $\sim 4 \times 10^{25}$~s for $\mx = 100$~MeV.  This bound becomes
lower by an order of magnitude at $\mx = 1$~TeV due to inefficient
energy deposition into the intergalactic medium.  We also show that
structure formation can enhance the effect of p--wave suppressed 
annihilation cross sections by many orders of magnitude with respect
to the background cosmological rate, although even with this
enhancement, CMB constraints are not yet strong enough to reach the
thermal relic value of the cross section.  
\end{abstract}

\maketitle

\section{Introduction}

The temperature and polarization fluctuations of the cosmic microwave
background (CMB) are sensitive to all redshifts since recombination,
and the large correlation between temperature and polarization at low
multipoles suggests rescattering of CMB photons at low $z$. Much of
this can be attributed to reionization by stars, but extra energy
injection into the intergalactic medium (IGM) at late times can
increase correlations on large scales.  As is well known, observations
of the CMB set severe constraints on weakly interacting massive
particles (WIMPs) models for dark matter (DM) candidates with masses
in the GeV mass range and below~\cite{Chen:2003gz, Hansen:2003yj,
  Pierpaoli:2003rz, Padmanabhan:2005es, Mapelli:2006ej, Zhang:2006fr,
  Ripamonti:2006gq, Zhang:2007zzh, Chuzhoy:2007fg, Finkbeiner:2008gw,
  Natarajan:2008pk, Natarajan:2009bm, Belikov:2009qxs, Galli:2009zc,
  Slatyer:2009yq, Cirelli:2009bb, Kanzaki:2009hf, Chluba:2009uv,
  Valdes:2009cq, Natarajan:2010dc, Hutsi:2011vx, Galli:2011rz,
  Finkbeiner:2011dx, Natarajan:2012ry, Evoli:2012zz, Giesen:2012rp,
  Evoli:2012qh, Slatyer:2012yq, Frey:2013wh, Cline:2013fm,
  Weniger:2013hja, Dvorkin:2013cga, Lopez-Honorez:2013cua,
  Ade:2013zuv, Madhavacheril:2013cna}.  In most of the studies on DM
annihilations, the CMB constraints have been typically derived
assuming that the annihilation cross section times the relative
velocity, $\sv$, is constant, \textit{i.e.}, s--wave annihilations (see
Refs.~\cite{Lopez-Honorez:2013cua, Ade:2013zuv} for the latest
results).  In this case, CMB mostly constrains the new sources of
ionization and heating due to the products of annihilations of the
homogeneous background DM distribution around the epoch of
recombination.  Furthermore, in the framework of constant $\sv$ and 
when halo formation based on N--body simulations is considered, the
CMB bounds are not influenced by late--time effects such as DM
clustering in structures~\cite{Lopez-Honorez:2013cua} (see also
Refs.~\cite{Natarajan:2008pk, Natarajan:2009bm, Belikov:2009qxs,
  Cirelli:2009bb, Kanzaki:2009hf, Natarajan:2010dc, Hutsi:2011vx,
  Natarajan:2012ry, Giesen:2012rp} for other results). 

In this paper, we turn our attention to two DM scenarios whose effect
on the CMB is expected to be driven by late--time ($z \lesssim 50$)
physics. We first revisit and update the case of DM
decay~\cite{Chen:2003gz,Cline:2013fm} assuming that it accounts for
the entire DM relic abundance.  Hence, the DM lifetimes $\tau_\chi$
considered here are assumed to be larger than the age of the Universe.
The second scenario studied in this paper is the case of DM species
with a velocity--suppressed annihilation cross section, specifically
$\sv \simeq b v^2$.  In the non--relativistic limit, the cross section
may be expanded as $\sv=a+b v^2+{\cal O} (v^4)$, with constant $a$ and
$b$ which govern the s--wave and p--wave contributions, respectively.
This gives rise to an averaged annihilation cross section times the
relative velocity $\sva$ and a DM relic abundance $\Omega_{\rm DM}h^2$
of the following form~\cite{Srednicki:1988ce, Gondolo:1990dk,
  Griest:1990kh}:   
\begin{equation}
\sva_{\rm FO} = a + 6b/x_{\rm FO}\quad \mbox{and} \quad   
\left(\frac{\Omega_{\rm DM} h^2}{0.1}\right) \simeq 0.34 \,
\left(\frac{x_{\rm FO}}{\sqrt{g_*}}\right) \, \left(\frac{3 \times
  10^{-26} {\mathrm{cm}}^3/{\mathrm s}}{a+3b/x_{\rm FO}}\right) ~,
\label{eq:relic}
 \end{equation}
when limiting the expansion in velocity to the $v^2$ contribution.  In
Eq.~(\ref{eq:relic}), $x_{\rm FO} \equiv \mx/\Tcd$, where $\mx$ is the
DM mass and $\Tcd$ is the temperature at which freeze--out (or chemical
decoupling) occurs, $g_*$ refers to the number of relativistic degrees
of freedom at the time of freeze--out. Eq.~(\ref{eq:relic}) implies
that s--wave annihilating DM requires $\sva_{\rm FO} \sim 3\times
10^{-26}$~cm$^3$/s in order to account for the correct cosmological  
abundance (see however Ref.~\cite{Steigman:2012nb} for an accurate
calculation), while DM annihilating through a p--wave channel requires
$\sva_{\rm FO} \sim 6\times 10^{-26}$~cm$^3$/s.   

The rate at which DM decays or annihilations heat or ionize the
baryonic component of the IGM is proportional to (see, \textit{e.g.},
Ref.~\cite{Padmanabhan:2005es})   
\begin{equation}
  {\cal F}(z)= \frac{1}{H(z)(1+z) n_{H}(z)}\left(\frac{\ud E}{\ud t
    \ud V}\right)_{\rm deposited} 
\label{eq:cFz}
\end{equation}
where $H(z)$ is the Hubble rate and $n_H(z)\propto (1+z)^3$ is the
density of hydrogen nuclei.  In the matter dominated era, the 
denominator $H(z) (1+z) n_{H}(z)$ suppresses energy deposition by a
factor of $(1+z)^{-11/2}$.  For the background DM component, the energy
deposition rate is usually expressed through 
\begin{equation}\label{rate}
 \left(\frac{\ud E}{\ud t \ud V}\right)_{\rm deposited} =
 f(z)\left(\frac{\ud E}{\ud t \ud V}\right)_{\rm injected} ~,
\end{equation}
where the factor $f(z)$, defined as
\begin{equation}
f(z) \equiv \frac{\mathrm{Energy \,  deposited \,  into \, IGM \, at
    \,} z}{\mathrm{Energy \, injected \, at \,} z} ~, 
\label{eq:fz}
\end{equation}
accounts for the fact that final--state energy can stream away with
neutrinos and that energy losses by final--state electrons, positrons
and photons occurs via a cascade of collisions, so they may be
absorbed at later times or freely stream until the present, or until
they are redshifted into a window in which the IGM is less
transparent.  We will return to the form of $f(z)$ in the different 
scenarios in Sec.~\ref{sec:models-energy-inject}. 
  
Prior to structure formation, the smooth dark matter background
contribution drives the energy injection rate $\left(\ud E/\ud t \ud V
\right)_{\rm injected}$. The latter is proportional to
\begin{equation}
  \Gamma_{\rm dec}=\frac{n_\chi}{\tau_\chi}
\label{eq:Gd}
\end{equation}
for DM decays. This scales as $(1+z)^3$, where $n_\chi=\rho_\chi/m_\chi$ 
the DM number density, while for DM annihilation the injection rate is
proportional to
\begin{equation}
  \Gamma_{\rm ann}=n_\chi^2\sva,
\label{eq:Ga}
\end{equation}
which scales as $(1+z)^6$ for an s--wave dominated cross section. As
a result, the redshift dependence of ${\cal F}(z)$ goes as
$(1+z)^{1/2}$ for s--wave DM annihilation cross sections whereas for
DM decays, it goes as $(1+z)^{-5/2}$.  This explains why DM decays
are expected to affect CMB at later times than s--wave DM
annihilations.  In the case of p--wave suppressed DM annihilation
cross sections, the two extra powers of $v$ in $\sva$ redshift with
time as $(1+z)^2$ so that the background injected energy rate goes as
$\Gamma_{\rm ann}\propto (1+z)^8/(1+z_{\rm KD})^2$, where $z_{\rm KD}$
is related to the time of DM kinetic decoupling --- when local
thermal equilibrium is not maintained any more by scattering with
Standard Model (SM) particles --- and which typically occurs well
before recombination $z_{\rm KD} \gg z_{\rm rec}$.  This will be
explored in greater detail in Sec.~\ref{sec:vdep}.  Although such a 
dependence appears to severely suppress the impact of p--wave DM
annihilations on CMB photons at all times, this discussion only
applies for the background contribution.  In contrast, we will show 
that the enhancement at late times of the $n_\chi^2$ and $v^2$
factors provided by the formation of DM halos dominates by many
orders of magnitude over the background contribution.  In both the
case of DM decays and p--wave DM annihilations, late--time
contributions to rescattering of CMB photons would mainly manifest as
a modification to the low--multipole polarization spectrum of the CMB,
albeit for different physical reasons.

In Sec.~\ref{sec:models-energy-inject}, after briefly characterizing
the time--dependence of energy injection into the IGM in general, we
further describe our treatment of energy injection from DM decay
(Sec.~\ref{sec:decaying-dark-matter}) and p--wave annihilation
(Sec.~\ref{sec:vdep}).  In order to obtain the limits for both scenarios
we use the latest available CMB data, including the recent Planck
data~\cite{Ade:2013zuv}, the nine--year temperature and polarization
data release from the Wilkinson Microwave Anisotropy Probe (WMAP)
collaboration~\cite{Hinshaw:2012aka} and the high--multipole CMB data
released by the South Pole Telescope
(SPT)~\cite{Story:2012wx, Hou:2012xq} and by the Atacama Cosmology
Telescope (ACT)~\cite{Sievers:2013ica} experiments.  We also add a
prior on the Hubble constant, $H_0$, from the Hubble Space Telescope
(HST)~\cite{Riess:2011yx} and from Baryon Acoustic Oscillation (BAO)
measurements from different surveys~\cite{Anderson:2012sa,
  Padmanabhan:2012hf, Beutler:2011hx, Blake:2011en}.  Finally, we also
use the determination of the IGM temperature from Lyman--$\alpha$
observations~\cite{Schaye:1999vr}, which significantly improves the
limits.  Our results are presented and discussed in
Sec.~\ref{sec:results} and we conclude in Sec.~\ref{sec:conclusions}.

\section{Energy injection scenarios}
\label{sec:models-energy-inject}

Final--state energy produced by DM decays or annihilations can stream
away in the form of neutrinos, while the daughter photons, electrons
and positrons may be absorbed at later times into the IGM via
different processes, which include photoionization, Coulomb
scattering, Compton processes, bremsstrahlung and recombination.
These effects can be written in terms of a transfer function
$T_i(z',z,E)$ for each channel $i = \{e^\pm,\gamma\}$, which describes
the fraction of the original particle's energy deposited into the IGM
at a redshift $z$, for a redshift of injection $z'$ and an initial
energy $E$, per logarithmic redshift bin $d\ln (1+z)$.  In this case,
``deposition'' simply means that the particles hit some threshold
energy, below which their absorption into the IGM as heat or
ionization energy can be considered instantaneous.

This has been computed by several authors~\cite{ Slatyer:2009yq,
  Valdes:2009cq, Evoli:2012zz} with Monte Carlo codes which track the
evolution of ``primary'' particles and of their ``secondary''
daughters in an expanding Universe.  Tabulated $T_i(z',z,E)$ have been
made
public\footnote{\texttt{http://nebel.rc.fas.harvard.edu/epsilon/}} by
Ref.~\cite{Slatyer:2012yq}.  Eq.~(\ref{eq:fz}) corresponds to a
normalized integral over all previously injected energy, which can be
written in a general form valid for different scenarios, as  
\begin{equation}
\label{fofz} 
 f_\alpha(z, \mx) =  \frac{\sum_i \int E\, \ud E \int \ud
  z' \, T_i(z',z,E) \, \frac{\ud N_i(E,\mx)}{\ud E
   \ud z'}}{\sum_i \int E\, \ud E \, \frac{\ud N_i(E,\mx)}{\ud E \ud
     \ln{(1+z)}}} ~, 
\end{equation}
where $\frac{\ud N_i(E,m_\chi)}{\ud E \ud \ln (1+z)}$ is the spectrum of
injected particles, per comoving volume, as a function of energy and
redshift for a given DM mass $\mx$.  Omitting the redshift--independent
and energy--independent proportionality factors, we have:
\begin{equation}
\frac{\ud N_i(E,m_\chi)}{\ud E \ud \ln (1+z)}\propto
\frac{\Gamma_\alpha(z)}{(1+z)^3H(z)} \frac{\ud N_i (E, \mx)}{\ud E} ~,
\end{equation}
where $\frac{\ud N_i (E, \mx)}{\ud E}$, is the energy spectrum of
final--state particles per DM annihilation or decay and
$\Gamma_\alpha$ is defined in Eqs.~(\ref{eq:Gd}) and~(\ref{eq:Ga}) and 
scales as $(1+z)^3, (1+z)^6$ and $(1+z)^8$ for decaying, s--wave and
p--wave annihilating dark matter species, respectively.  The energy
spectrum of decays and annihilations into $e^+e^-$ and $\mu^+\mu^-$ is
computed as in Ref.~\cite{Lopez-Honorez:2013cua} and for the 
$\tau^+\tau^-$ channel we use the publicly available
results\footnote{\url{http://www.marcocirelli.net/PPPC4DMID.html}} 
of Refs.~\cite{Ciafaloni:2010ti, Cirelli:2010xx}.  

In the present work, we use the $T_i(z',z,E)$ functions obtained in
Ref.~\cite{Slatyer:2012yq}.  Note, however, that these results neglect
heating from proton and antiproton final--state particles, which may
account for up to 20\% of the deposited energy for some
channels~\cite{Weniger:2013hja}.  This effect would change the CMB
bounds on DM at the 10\% level.  It was also recently pointed out that
errors in the standard computation of $f(z)$ may introduce systematic
uncertainties that could weaken constraints \cite{Galli:2013dna},
although in practice the effects of these on energy injection bounds
turn out to be small~\cite{Madhavacheril:2013cna}.

\subsection{DM decays}
\label{sec:decaying-dark-matter}

In general, the decay of unstable particles may affect the redshift of
recombination, as well as reionization at low redshift (see
\textit{e.g.}, Refs.~\cite{Doroshkevich:2002ff, Bean:2003kd,
  Chen:2003gz, Mapelli:2006ej, Ripamonti:2006gq, Zhang:2007zzh,
  Slatyer:2012yq, Finkbeiner:2011dx, Cline:2013fm}).  The rate of
energy per unit volume liberated by such an unstable cosmological
species $\chi$ reads  
\begin{equation}
\label{injection}
 \left(\frac{\ud E}{\ud t \ud V}\right)_{\rm injected} = (1+z)^3\,
 \frac{\phi_{\rm SM}}{\tau_\chi} \, \, \Omega_\chi \rho_c ~, 
\end{equation}
and depends linearly on its number density, $n_\chi = \Omega_\chi
\rho_c (1+z)^3/m_\chi$, where $\Omega_\chi$ is the contribution of
$\chi$ to the critical density $\rho_c$, and $\phi_{\rm SM}$ is the
fraction of its mass that goes to into SM particles when decaying (while
$(1-\phi_{\rm SM})$ would go to some other dark sector species).  Here
we assume that $\chi$ accounts for all the DM such that $\Omega_\chi =
\Omega_{\rm DM, 0}$, it fully decays to SM particles ($\phi_{\rm
  SM}=1$) and has a lifetime $\tau_\chi \gg t_{\rm U}$ where $t_U =
4.34 \times 10^{17}$~s is the age of the Universe.  To consider
particles with significantly shorter lifetimes, one would include an
exponential factor in (\ref{injection}) to parametrize the species'
depletion.

In practice, we have used \texttt{CosmoRec}
package~\cite{Chluba:2010ca, Chluba:2009uv, AliHaimoud:2010ab,
  Chluba:2010fy, Grin:2009ik, Switzer:2007sq, RubinoMartin:2009ry} in
order to compute the changes in the ionization history.
\texttt{CosmoRec} includes a subroutine that modifies the evolution
equations for the IGM temperature and for the net ionization rate from
the ground states of neutral hydrogen and helium, which depends on the
energy deposition by the DM.  In the case of DM decays, inspired by
Ref.~\cite{Chluba:2009uv}, we use the parametrization
\begin{equation}
 \left(\frac{\ud E}{\ud t \ud V}\right)_{\rm deposited} = f_{\rm eff,
   dec}(m_\chi) \, \epsilon_{\rm dec} \, n_H(z) \, \,
      {\mathrm{eV/s}}\quad {\mathrm{with}}\quad \epsilon_{\rm dec} =
      7.2 \times 10^{-14} \,
      \left[\frac{10^{23}\,{\mathrm{s}}}{\tau_\chi}\right]\,
      \left[\frac{\Omega_{\rm DM, 0} h^2}{0.13}\right] ~, 
\label{eq:EdecCR}
\end{equation}
where $n_H (z) = 1.9 \times 10^{-7} \, \mathrm{cm}^{-3} \, (1+z)^3$ is
approximately the number density of hydrogen nuclei in the Universe.

Analogously to Ref.~\cite{Lopez-Honorez:2013cua}, we define an effective
$f_{\rm dec}(z,m_\chi)$ averaged over redshift which depends on the DM
mass only and allows us to simplify the numerical analysis.  This
allows for constraints computed with a given final state and
$z$--dependence to be rescaled for each value of the DM mass.
However, the redshift dependence of $f_{\rm dec}(z, \mx)$ in the case
of decays is much more pronounced than for annihilations and, for
masses above $\sim$1~GeV, it is not possible to accurately reproduce
the free electron fraction $x_e(z)$ with a constant $f_{\rm eff}(\mx)$
for each DM mass, in turn causing effects at intermediate redshifts to
be underestimated when using the parametrization in
Ref.~\cite{Lopez-Honorez:2013cua}.  Nevertheless, the injection of
additional energy broadens the last scattering surface by increasing
the residual ionization, without slowing
recombination~\cite{Padmanabhan:2005es}.  This is reflected in the
visibility function, $\tilde g(z)= - \tau' e^{-\tau}$, where $\tau$ is
the optical depth and $'$ indicates the derivative with respect to the
conformal time, $\eta$.  Whereas around recombination, the visibility
function is approximately the same with or without the small energy
injection from DM, below $z_{\rm max} \simeq 600-800$ (depending on
the decay channel and lifetime), energy injection creates a longer
tail on $\tilde g(z)$.  At these times, the optical depth is very
small, so the exponential factor in $\tilde g(z)$ is approximately
one, and the visibility function is $\tilde g(z) \simeq -\tau' \propto
x_e(z) (1+z)^2$.  By using the redshift dependence of
$x_e(z)$~\cite{Lopez-Honorez:2013cua}, we can define $\tilde g_{\rm
  eff}$ with $f_{\rm eff, dec}(\mx)$ determined by imposing
$\int_{\eta({\rm z_{max}})}^{\eta(0)} \ud \eta \, \tilde g(\eta) =
\int_{\eta({\rm z_{max}})}^{\eta(0)} \ud \eta \, \tilde g_{\rm
  eff}(\eta)$, which reads
\begin{equation}
f_{\rm eff, dec}(\mx) = \frac{\int_{z_{\rm max}}^0
  \frac{(1+z)^2}{H(z)} \, \ud z \, \int_{\infty}^{z} 
  \frac{\Gamma_{\rm dec}(z')}{(1+z')^4H(z')} \, f_{\rm dec}(z',\mx) \,
  \ud z'} {\int_{z_{\rm max}}^0 \frac{(1+z)^2}{H(z)} \, \ud z \,
  \int_{\infty}^{z} \frac{\Gamma_{\rm dec}(z')}{(1+z')^4H(z')} \, \ud
  z'} ~, 
\label{eq:feffdecays}
\end{equation}
where $\Gamma_{\rm dec}$ is defined in Eq.~(\ref{eq:Gd}).  We have
checked that for $\chi \rightarrow e^+e^-$, $f_{\rm eff, dec}(\mx)$
leads to constraints within 5\% of those computed using the full
$f_{\rm dec}(z,\mx)$, for values of $\mx$ sampled in the full range
considered, between $\sim$2~MeV and 1~TeV, using $z_{\rm max} = 800$.
For decays to $\mu^+\mu^-$ and $\tau^+\tau^-$, constraints obtained
with Eq.~(\ref{eq:feffdecays}) are within 20\% and 15\% of those
obtained with $f(z,\mx)$, respectively. In both cases we use $z_{\rm
  max} = 600$.  Tabulated values of $f_{\rm eff, dec}(\mx)$ for 
these three decay channels are given in
Appendix~\ref{feffappendix}\footnote{Let us note that the quoted
  values for $z_{\rm max}$ are not obtained from a fit to $f_{\rm
    dec}(z,\mx)$, but are educated choices which provide accurate
  results.}.

\begin{figure}[t]
\begin{tabular}{c c}
\includegraphics[width=0.5\textwidth]{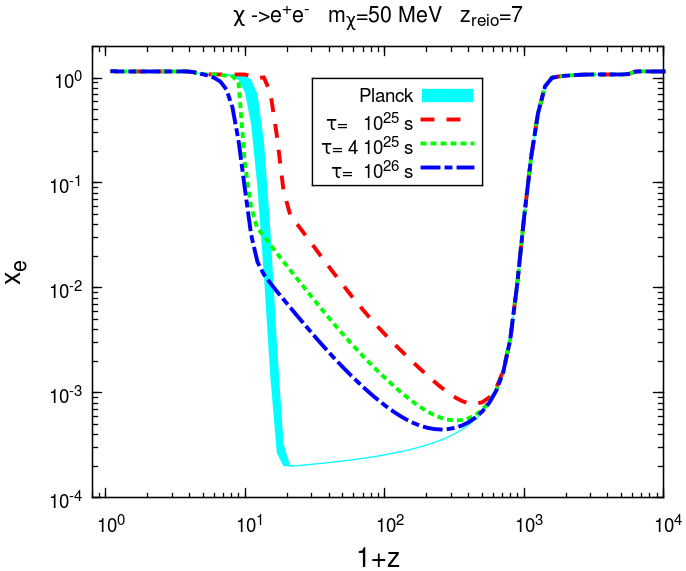}
&
\includegraphics[width=0.5\textwidth]{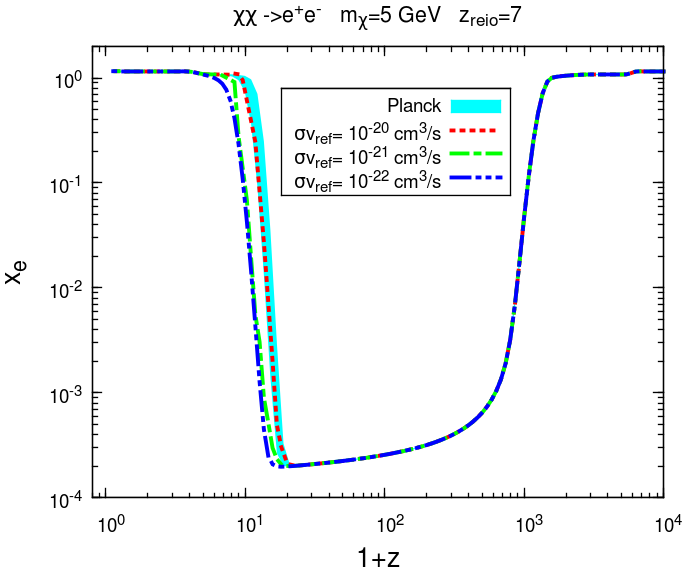}
\\ 
\end{tabular}
\caption{\textit{The free electron fraction $x_e$ as a function of
    redshift $1+z$ in the case of DM decays (left) and p--wave
    annihilations (right).  Both scenarios can enhance ionization at
    late times, due to the redshift dependence of their energy
    release.}}  
\label{fig:ion}
\end{figure}

In Fig.~\ref{fig:ion}, we show the resulting free electron fraction
$x_e(z)$ as a function of the redshift from the recombination period
with and without extra energy injection by DM.  The light blue band is
shown for reference and corresponds to the Planck 95\% confidence
level (CL) determination\footnote{
  \url{http://www.sciops.esa.int/SYS/WIKI/uploads/Planck_Public_PLA/3/32/Grid_limit95.pdf}   
baseline model $2.1$} of the optical depth to reionization $\tau_{\rm
  reio}= 0.089^{+0.027}_{-0.024}$ (in the absence of a DM
contribution).  In the left panel of Fig.~\ref{fig:ion}, the three 
other curves illustrate the impact on $x_e(z)$ of a 50~MeV DM 
candidate decaying 100\% into electron/positron pairs with three 
different lifetimes.  In addition to the DM contribution, a simplified
model for reionization from stars at $z_{\rm reio}=7$, as implemented
in the CAMB code~\cite{Lewis:1999bs}, has been taken into account in
each case (in the standard case in which reionization is induced by
star formation only, current cosmological measurements indicate
$z_{\rm reio}\sim 11$~\cite{Ade:2013zuv}).  However, let us note that
for our Monte--Carlo--generated constraints, we leave the redshift of
reionization as a free parameter, as indicated below.  As expected,
the effect of long--lived decaying DM becomes important at late times
and, as we shall illustrate, lifetimes such as $\tau_\chi = 10^{25}$~s
are clearly excluded by CMB data for a 50~MeV DM candidate decaying
into $e^+e^-$.

\subsection{DM p--wave annihilations}
\label{sec:vdep}

The second scenario that we examine involves DM annihilating
predominantly through p--wave processes, \textit{i.e.}, $\sv \simeq b
v^2$ with constant $b$.  In this case, the total deposited energy is
parametrized as 
\begin{equation}
\left(\frac{\ud E}{\ud V \ud t} \right)_{\rm deposited} = \left[
  \left( \frac{1+z}{1+\zref}\right)^2f_{\rm p}(z,\mx) + g_{\rm
    p}(z,\mx,\vref) \right] (1+z)^6 \, \rho_{\chi}^2 \,
\frac{\svref}{m_\chi} ~,  
\label{eq:injectedE}
\end{equation}
where the first term in the square brackets accounts for the
background DM contribution and the second term for the halo
contribution, with $\rho_{\chi} = \Omega_{\rm DM, 0} \rho_{c}$.  We
have written the p--wave suppressed annihilation cross section $\sva$
as $\sva=\svref {\langle v^2\rangle}/{\vref^2}$, defining an arbitrary
reference velocity $\vref$.  The background component is proportional
to $f_{\rm p}(z,m_\chi)$, defined in Eq.~(\ref{fofz}) with
$\Gamma_\alpha$ of Eq.~(\ref{eq:Ga}) scaling as $(1+z)^8$.  We
made use of the fact that, after the time of kinetic decoupling, the
temperature of non--relativistic species in an expanding Universe goes
as $T\propto (1+z)^2$ so that, using equipartition of energy, $\langle
v^2\rangle$ is related to $\vref$ or equivalently $\zref$ (at which
$\svref$ is evaluated) through
\begin{equation}
 \frac{\langle v^2\rangle}{\vref^2} = \frac{T_{\chi}(z)}{T_{\rm ref}} =
 \left(\frac{1+z}{1+\zref} \right)^2 ~.
\label{redshiftevolution}
\end{equation}
In this work, we present our results for $\vref = 100$~km/s, since that
is the order of the dispersion velocity of DM in halos today and is
thus the relevant quantity for comparison with indirect DM searches.
In the case of the halo contribution, the relevant ${\langle v^2
  \rangle}/{\vref^2}$ factor has been absorbed into the definition of
$g_{\rm p}(z,\mx,\vref)$ and will be given explicitly in
Sec.~\ref{sec:halo-contribution}.

\subsubsection{Background DM contribution}
\label{sec:backgr-contr}

One can estimate $\zref$, the redshift at which the root mean square
velocity $v_{\rm rms} \equiv \sqrt{\langle v^2\rangle }$ of the
background DM is equal to $\vref$, as a function of the redshift of
kinetic decoupling $\zkd$ and the corresponding temperature $T_\chi
(\zkd) = T_{\rm KD}$.  Using Eq.~(\ref{redshiftevolution}) and
equipartition of energy for an ideal gas,   
\begin{equation}
  1+\zref = \frac{\vref}{c}\, (1+\zkd) \left(
  \frac{\mx}{3 \, \Tkd}\right)^{1/2}\,.
\label{eq:zref-zkd}
\end{equation}
Furthermore, one can express $\zkd$ in terms of $T_{\rm KD}$ and the
CMB temperature by using the fact that DM was in thermal equilibrium
with the CMB at the time of kinetic decoupling,
\begin{equation}
1+\zkd = \frac{T_{\rm KD}}{T_{\rm CMB,0}} \simeq 4.2 \times 10^{9}
\left(\frac{T_{\rm KD}}{\mathrm{MeV}}\right) ~,
\label{kdeq}
\end{equation}
where $T_{\rm CMB,0} = 0.238$~MeV is the temperature of the CMB today.
Combining Eqs.~(\ref{eq:zref-zkd}) and (\ref{kdeq}),
\begin{equation}
1+\zref \simeq 2.56 \times 10^{7} \left(\frac{\Tkd}{\mathrm{MeV}}
\right)^{1/2} \left(\frac{\mx}{\mathrm{GeV}}\right)^{1/2}.
\label{eq:zref}
\end{equation}

The temperature of kinetic decoupling is model--dependent and has been
computed by several authors on a case--by--case
basis~\cite{Schmid:1998mx, Zybin:1999ic, Chen:2001jz, Hofmann:2001bi,
  Berezinsky:2003vn, Green:2005fa, Loeb:2005pm, Profumo:2006bv,
  Bertschinger:2006nq, Bringmann:2006mu, Bringmann:2009vf,
  Cornell:2012tb, Gondolo:2012vh, Cornell:2013rza, Shoemaker:2013tda}.
Of interest for the annihilating DM scenario considered here,
Ref.~\cite{Shoemaker:2013tda} considered fermionic DM candidates 
annihilating into SM leptons through effective interactions suppressed 
by an energy scale $\Lambda$ that give rise to p--wave suppressed
annihilation cross section for scalar type interactions.  In this
case, assuming that the DM mass is much larger than the final--state
lepton mass, Ref.~\cite{Shoemaker:2013tda} obtained a temperature of
kinetic decoupling
\begin{equation}
\Tkd = 0.69 \, \frac{g_{\rm eff}^{1/8}}{g_\chi^{1/4}} \, \Lambda \, 
\left(\frac{48\pi \, m_\chi}{M_{pl}}\right)^{1/4}  \simeq 2.02 \,
\mathrm{MeV} \, \left(\frac{m_\chi}{\mathrm{GeV}}\right)^{3/4} ~.
\label{eq:TKD-pwav}
\end{equation}
For the second equality above, we have taken $g_{\rm eff} \simeq 100$
(relativistic degrees of freedom at $T_{\rm KD}$), $g_\chi = 2$ (internal
degrees of freedom of the DM particle) and $\Lambda$ was chosen so
that the annihilation cross section at the time of freeze--out matches
$\sva = 6 \times 10^{-26}$~cm$^3$~s$^{-1}$. The value of $\Tkd$
obtained by Ref.~\cite{Shoemaker:2013tda} considering the case of
vector type interactions giving rise to an s--wave annihilation cross
section only differs from the p--wave case by a few percent.  As an
additional example, the kinetic decoupling temperature for neutralino
DM from Ref.~\cite{Chen:2001jz} is estimated to be 
\begin{equation}
\Tkd \sim \mathrm{MeV} \left(\frac{\mx}{\mathrm{GeV}}\right)^{2/3}.
\end{equation}

Thus, in general, kinetic decoupling occurs at a later stage than
chemical freeze--out, which takes place at $T_{\rm FO} \simeq \mx/20$.
Combining Eq.~(\ref{eq:TKD-pwav}) with Eq.~(\ref{eq:zref}), one can
see that the background contribution in Eq.~(\ref{eq:injectedE}) is
severely suppressed at redshifts $z\sim 10^3$ and below, even for
$f_{\rm p}(z, \mx) = 1$.  In fact, the suppression of the background
contribution at the epoch of recombination, combined with the velocity
enhancement at late times in DM structures, makes the halo contribution
$g_{\rm p}(z,\mx,v_{\rm ref})$ in Eq.~(\ref{eq:injectedE}) dominate
the energy deposition history by many orders of magnitude, in contrast
to the case of s--wave annihilations.  This is discussed in the
following section.

\subsubsection{Halo contribution}
\label{sec:halo-contribution}

At late times, the formation of halos not only enhances the DM average
squared number density $\langle n_\chi^2 \rangle$, but also the
average of the square of the DM particles relative velocity, $\langle 
v^2 \rangle$.  This is simply due to a transfer of gravitational
potential energy into kinetic energy of the individual particles that
make up each halo.  In order to illustrate the importance of the halo
contribution, we define an effective DM density
\begin{equation}
\rho_{\rm eff, s} = \rho_{\chi} (1+z)^3 \left(1 + G_{\rm s}(z)
\right)^{1/2} ~, 
\label{rhoeffs}
\end{equation}
for s--wave dominated annihilation~\cite{Cirelli:2009bb}, and
analogously 
\begin{equation}
\rho_{\rm eff,p} = \rho_{\chi} (1+z)^3
\left(\left(\frac{1+z}{1+\zref}\right)^2 + G_{\rm p}(z,\vref)
\right)^{1/2} ~,  
\label{rhoeffp}
\end{equation}
for p--wave dominated annihilation.  

Both effective DM densities depend on a dimensionless halo
contribution defined as
\begin{equation}
G_{\rm s}(z) \equiv \frac{1}{\rho_{\chi}^2} \, \frac{1}{(1+z)^6} \,
\int \ud M \, \frac{\ud n(M,z)}{\ud M}  \, \int_0^{r_{\Delta}} \ud r
\, 4 \pi r^2 \, \rho_{\rm halo}^2(r) ~, \label{eq:G}
\end{equation}
for s--wave annihilation~\cite{Lopez-Honorez:2013cua}, and as
\begin{equation}
G_{\rm p}(z,\vref) \equiv \frac{1}{\rho_{\chi}^2} \, \frac{1}{(1+z)^6}
\, \int \ud M \, \frac{\ud n(M,z)}{\ud M}  \, \int_0^{r_{\Delta}} \ud
r \, 4 \pi r^2 \, \frac{\langle v^2(r) \rangle}{\vref^2}\rho_{\rm
  halo}^2(r) ~,  
\label{eq:Gv}
\end{equation}
for p--wave annihilation. 
In both cases, $\ud n(M,z)/\ud M$ is the halo mass function and
$\rho_{\rm halo}(r)$ is the density profile of each individual halo
with virial radius $r_{\Delta}$.  Here we use the results of N--body
simulations from Ref.~\cite{Watson:2012mt} for the halo mass function
and from Ref.~\cite{Prada:2011jf} to obtain the relation of the
concentration parameter to the halo mass assuming a Navarro, Frenk and
White (NFW) DM density profile~\cite{Navarro:1995iw} for each
individual halo.  In Eq.~(\ref{eq:Gv}), the extra factor of $\langle
v^2(r) \rangle/\vref^2$ accounts for the halo--dependent velocity
boost.  The angular brackets represent an average over the square of
the DM velocity distribution in the halo, which we take to follow a
Maxwell--Boltzmann distribution. 

\begin{figure} [t]
\includegraphics[width=0.8\textwidth]{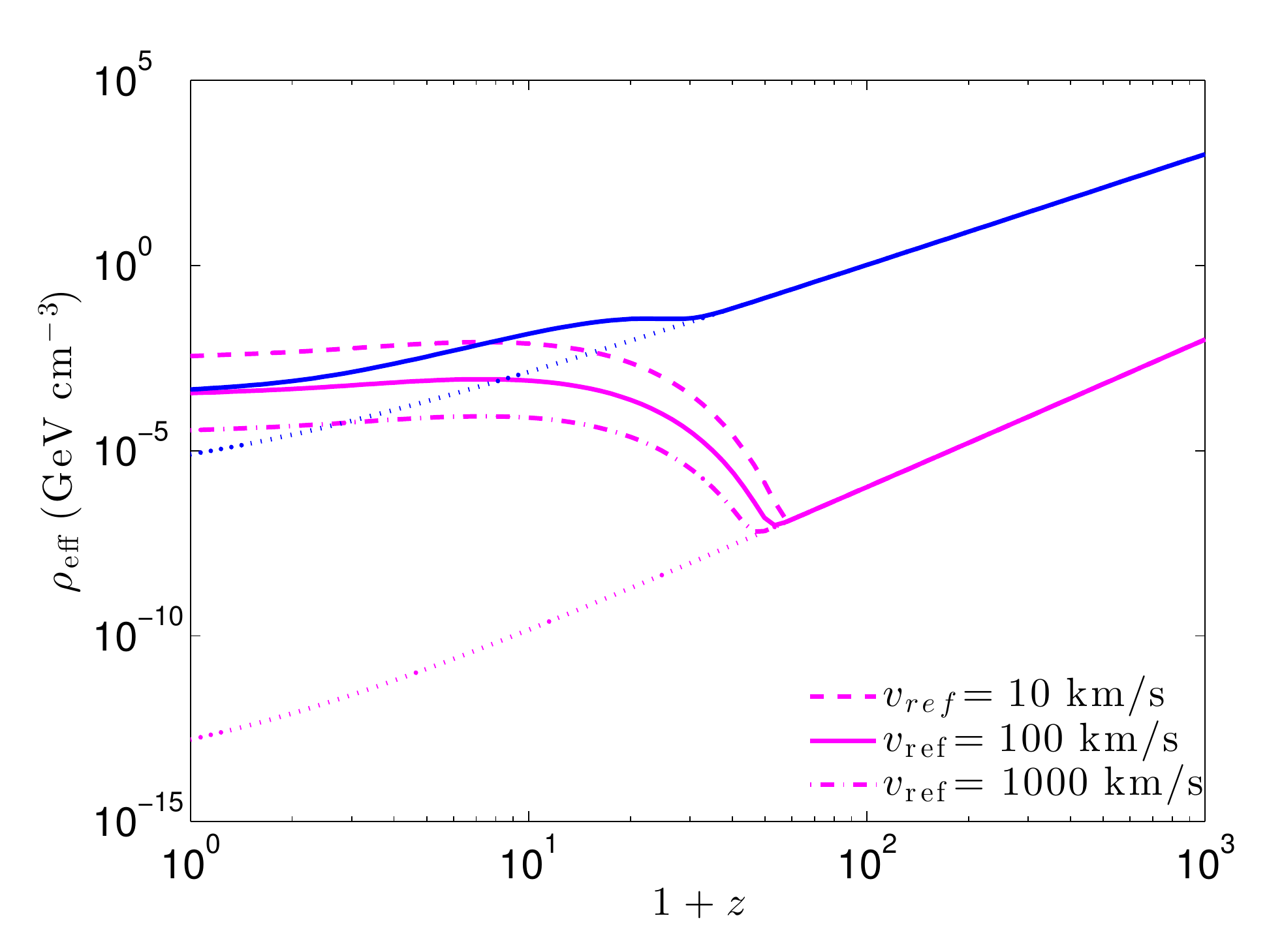}
\caption{\textit{Effective DM density $\rho_{\rm eff}$ as defined in
    Eqs.~(\ref{rhoeffs}) (blue upper curves: s--wave annihilation) and
    (\ref{rhoeffp}) (magenta lower curves: p--wave annihilation).  In
    each case, the monotonically increasing dotted line is the smooth
    DM background contribution, and the bump at low $z$ represents the
    halo contribution.  The p--wave case is given for three values of
    $\vref$ and we have chosen a redshift of thermal decoupling $\zkd =
    10^8$.  This figure is for illustration and does not represent a
    particular model of kinetic decoupling.}} 
\label{fig:rhoeff}
\end{figure}

The energy injected into the IGM at a given redshift from the
annihilation of DM particles, both by the background and the halo DM
contributions, depends on $\rho_{\rm eff}^2$, which is depicted in
Fig.~\ref{fig:rhoeff} for s--wave and p--wave annihilations.  This
figure clearly illustrates that, while the overall energy injected
from DM is smaller in the p--wave case, the relative contribution from
halos is much larger than in the s--wave case, providing a potentially
distinct imprint in the CMB power spectrum.  

Let us note that the changing fraction of the Universe's DM that is
contained in halos, $\phi_{\rm halo}(z)$, is defined as $\int \ud M M
\ud n/\ud M = \phi_{\rm halo}(z) \rho_\chi(z)$.  To account for this,
the first and second terms in parentheses in
Eqs. (\ref{rhoeffs}--\ref{rhoeffp}) should respectively be multiplied
by $(1 - \phi_{\rm halo})^2$ and $\phi_{\rm halo}$.  However, $G$ is a
rapidly growing function that is correlated with $\phi_{\rm halo}$.
This means that the interval during which both terms are important is
short and, to first approximation, this correction can be ignored. 

For the purpose of simplifying our numerical analysis, the total halo
contribution in Eq. (\ref{eq:injectedE}), including energy deposition
efficiency effects, can be approximated by
\begin{equation}
g_{\rm p}(z,\mx,\vref) \simeq f_{\rm s}(z,\mx) \, G_{\rm p}(z,\vref)
\label{gapprox}
\end{equation}
and 
\begin{equation}
G_{\rm p}(z,\vref) = \left(\frac{100 \mathrm{\, km /s}}{\vref}\right)^2
G_{\rm p}(z,100 \mathrm{\, km /s}) ~. 
\end{equation}
In Appendix~\ref{haloappendix} we provide the steps to compute the
complete halo contribution, along with a fitting formula that
approximates this quantity, which we use to compute the exclusion
regions presented in Section~\ref{sec:results}. Finally, we note that this approach does not include the additional effect of substructure, which could serve to boost the late-time signal from annihilating DM -- and thus tighten constraints -- even further. 

Since the background contribution to ionization by p--wave DM
annihilations is highly suppressed, the halo contribution, which
begins with structure formation around $z \sim 50$, would dominate any
observable effects.  In the right--hand panel of Fig.~\ref{fig:ion}, we
show the free electron fraction $x_e(z)$ as a function of the redshift
with and without DM contributions.  Again, the light blue band
corresponds to Planck results in the absence of a component from DM
annihilation.  The three other curves show $x_e(z)$ due to a 5~GeV DM
candidate which fully annihilates into $e^+e^-$ through a p--wave
suppressed process.  We simultaneously consider reionization by stars
at $z_{\rm reio} = 7$, which cannot account for Planck results by
itself.  In Fig.~\ref{fig:ion} we see that, for instance, if $\svref\sim
10^{-22}$~cm$^3$, a mixed DM--stars reionization scenario matches CMB
data (although a larger $z_{\rm reio}\sim 9$ is actually necessary to
get a good agreement with CMB data with such a cross section).

\begin{table}[t]
\begin{center}
\begin{tabular}{c|c}
\hline\hline
 Parameter & Prior\\
\hline
$\Omega_{b, 0} h^2$ & $0.005 \to 0.1$\\
$\Omega_{\rm DM, 0} h^2$ & $0.01 \to 0.99$\\
$\Theta_s$ & $0.5 \to 10$\\
$z_{\rm reio}$ & $7 \to 12$\\
$n_{s}$ & $0.5 \to 1.5$\\
$\ln{(10^{10} A_{s})}$ & $2.7 \to 4$\\
$\tau_\chi/(10^{24} \textrm{s})$ & $10^{-2} \to 10^{5}$\\
$\sigma \vref / (3 \times 10^{-26}
\textrm{cm}^3/\textrm{s})$ 
&  $10^{0} \to 10^{12}$\\ 
\hline\hline
\end{tabular}
\caption{\sl Priors on the cosmological parameters used in this work,
  including the DM lifetime and the p--wave annihilation cross
  section.}    
\label{tab:priors}
\end{center}
\end{table}

\section{Results}
\label{sec:results}

In this section we present the results of our Markov Chain Monte Carlo
(MCMC) studies for both DM decays and p--wave annihilations.  For each
DM mass $\mx$, the parameters considered in these two MCMC analyses
are   
\begin{equation}
\label{eq:parameters}
  \{\omega_b,\omega_{\rm DM}, \Theta_s, z_{\rm reio}, n_s,
  \log[10^{10}A_{s}]\} \; \; \; + \; \tau_\chi \;
  ({\mathrm{decays}}) 
  \; \;  {\mathrm{or}} \; \; \sigma \vref \; ({\mathrm{p--wave \;
    annihilations}}) ~, 
\end{equation}
where $\omega_b \equiv \Omega_{b, 0} h^{2}$ and $\omega_{\rm DM} \equiv
\Omega_{\rm DM, 0}h^{2}$ are the physical baryon and cold DM energy
densities today, $\Theta_{s}$ is the ratio between the sound horizon
and the angular diameter distance at decoupling, $z_{\rm reio}$ is the
reionization redshift, $n_s$ is the scalar spectral index, $A_{s}$ is 
the amplitude of the primordial spectrum, and $\tau_\chi $ and $\sigma
\vref$ are the DM lifetime and the p--wave cross section times the
relative velocity (for a reference value $\vref$), respectively.  We
make use of the Boltzmann code CAMB~\cite{Lewis:1999bs} as well as the
publicly available MCMC package \texttt{cosmomc}~\cite{Lewis:2002ah}
with the recombination module \texttt{CosmoRec}~\cite{Chluba:2010ca,
  Chluba:2009uv, AliHaimoud:2010ab, Chluba:2010fy, Grin:2009ik,
  Switzer:2007sq, RubinoMartin:2009ry}.  We show in
Tab.~\ref{tab:priors} the flat priors on the above parameters.

We have performed an analysis with the WMAP9 data~\cite{Hinshaw:2012aka}
(temperature and polarization) combined with SPT
data~\cite{Story:2012wx, Hou:2012xq}, which includes nuisance
parameters related to the Sunyaev--Zeldovich amplitude, $A_{SZ}$, to
the amplitude of the clustered point source contribution $A_C$, and to
the amplitude of the Poisson distributed point source contribution 
$A_P$.  We have also separately included high multipole data from the
ACT CMB experiment~\cite{Sievers:2013ica}, obtaining very similar
constraints to the case with SPT, which we do not show.  In addition
to CMB measurements, we include a prior on the Hubble constant $H_0$
from the HST~\cite{Riess:2011yx} and BAO measurements from a number of
surveys~\cite{Anderson:2012sa, Padmanabhan:2012hf, Beutler:2011hx,
  Blake:2011en}.  Nevertheless, the addition of the former two
external data sets does not significantly improve the results.  We
have also performed an analysis with the recent Planck CMB
data~\cite{Ade:2013zuv}, considering the high--$\ell$ TT likelihood
with measurements up to $\ell_{\rm max}=2500$, combined with the 
low--$\ell$ TT likelihood, which accounts for measurements up to
$\ell=49$ and the low--$\ell$ ($\ell=23$) TE, EE, BB
likelihood~\cite{Bennett:2012zja} by including WMAP9 polarization
measurements.  We include the lensing likelihood as well as external
data from HST and BAO measurements.  High multipole information from
both ACT and SPT experiments is also added, this time simultaneously,
following the analyses presented by the Planck collaboration.  All
foreground parameters have been marginalized over as in
Ref.~\cite{Ade:2013zuv}.

Following previous works~\cite{Cirelli:2009bb, Giesen:2012rp,
  Lopez-Honorez:2013cua}, we have also considered the IGM temperature
as an additional constraint.  Lyman--$\alpha$ observations indicate
that the IGM temperature is of the order of a few times $10^{4}$~K in
the redshift interval $2<z<4.5$~\cite{Schaye:1999vr}. Thus, the total
likelihood is supplemented by the temperature likelihood, by means of
a half-gaussian distribution with a mean $T_{\rm m}=11220$~K and a
standard deviation $\sigma_{T_{\rm m}}=8780$~K at a redshift $z=4.3$.
In other words, we only consider temperature bounds when the IGM
temperature of a given model at $z=4.3$ is larger than $T_{\rm
  m}=11220$~K. 

\begin{figure}[t]
\includegraphics[width=0.8\textwidth]{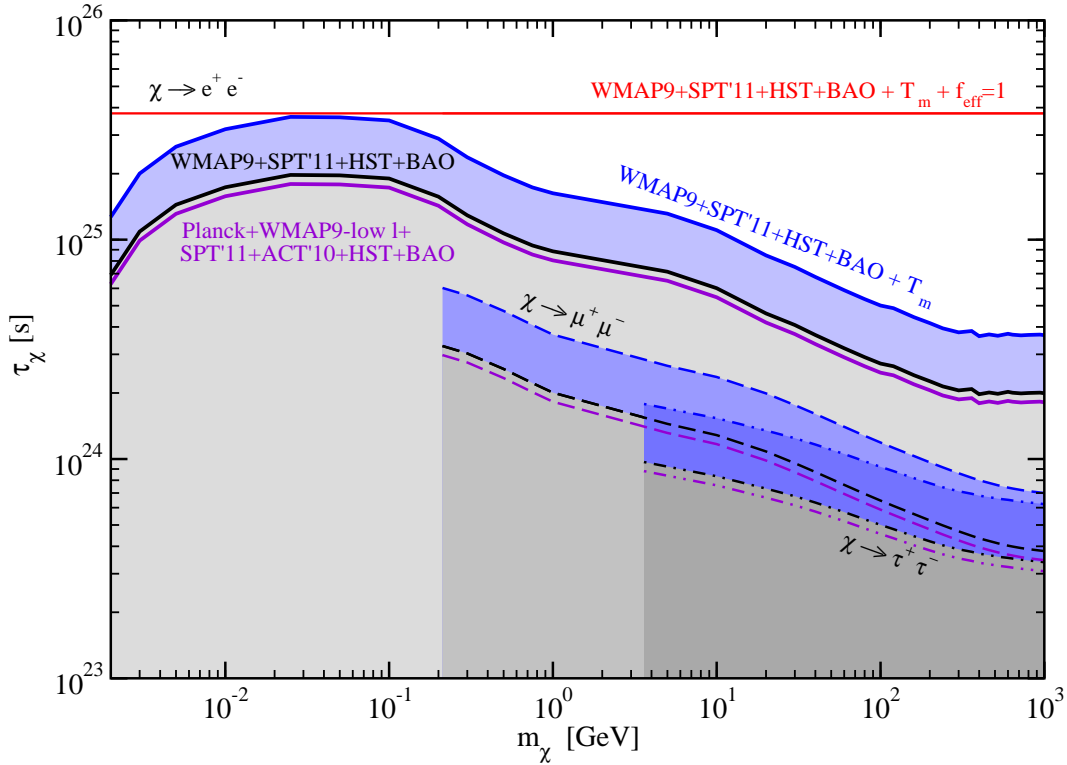}
\caption{\textit{Limits on the DM lifetime $\tau_\chi$ at 95\%~CL.  The
    upper red horizontal line assumes $100\%$ energy deposition
    efficiency.  Notice that Planck and WMAP9 constraints (combined
    with the external data sets) are very similar.  Including the
    efficiency $f_{\rm eff, dec}(\mx)$ gives the weaker, colored
    constraints.  We illustrate decays into $e^+e^-$ (longest solid
    lines), $\mu^+\mu^-$ (long dashed lines) and $\tau^+\tau^-$ (short
    dashed--dotted lines).  The different colors refer to the
    different data sets: black (blue) and violet refer to WMAP9 plus
    SPT'11 plus HST+BAO (plus also the prior on $T_{\rm m}$) and to 
    Planck plus WMAP9 low--$\ell$ polarization measurements plus CMB
    high--$\ell$ (ACT'10 and SPT'11) plus HST+BAO data sets,
    respectively.  Constraints on decays into two quarks or two weak
    gauge bosons lie between the $e^+e^-$ and $\tau^+\tau^-$ lines.}}  
\label{fig:decay}
\end{figure}

In Fig.~\ref{fig:decay} we show the $95\%$~CL limits on DM decays in the
($m_\chi$, $\tau_\chi$) plane for three different channels, which
bracket the limits into other SM decay channels. Unlike what occurs in
the case of s--wave DM annihilations, the measurement of the IGM
temperature $T_{\rm m}$ contributes significantly to the bounds and
further constrains DM energy injection due to the redshift dependence
of DM decays, with late injection becoming increasingly important just
as bounds from the IGM temperature start to be significant.  The most
stringent lower limit on the DM lifetime we obtain is
$\tau_\chi/f_{\rm eff, dec}(\mx) \gtrsim 4 \times 10^{25}$~s.  Let us
mention that bounds from gamma--ray searches are at the level of
$\tau_\chi \gtrsim 10^{26}-10^{27}$~s, depending on the target region
and DM mass~\cite{Abdo:2010dk, Dugger:2010ys, Huang:2011xr,
  Cirelli:2012ut, Ackermann:2012rg}, and bounds from antiproton
searches on hadronically decaying DM are slightly
stronger~\cite{Garny:2012vt, Delahaye:2013yqa}.  On the other hand,
our limits are better than those from neutrino
searches~\cite{PalomaresRuiz:2007ry, Covi:2009xn, Abbasi:2011eq} for
$\mx \lesssim 100$~GeV.  See Ref.~\cite{Ibarra:2013cra} for a recent
update (and a more complete list of references) on limits for unstable
DM.

\begin{figure}[t]
\includegraphics[width=0.8\textwidth]{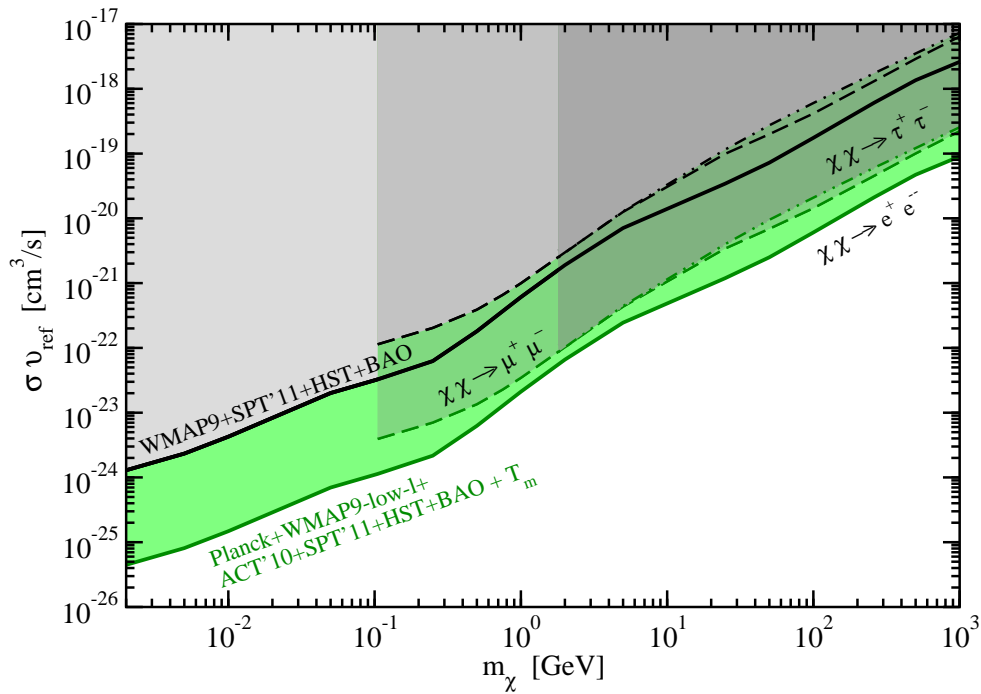}
\caption{\textit{Limits on the p--wave DM annihilation cross section
    $\sigma \vref=b \vref^2$ with $\vref =100$~km/s, at 95\%~CL.  The
    grey regions represent the constraints from CMB measurements only
    (WMAP9+SPT'11) plus HST+BAO, whereas the green regions use 
    Planck data, WMAP9 low--$\ell$ data, ACT'10+SPT'11, HST+BAO, as
    well as with a prior on the IGM temperature $T_{\rm m}$.  See the
    text for details.  The different lines represent DM annihilations
    into $e^+e^-$ (solid lines), $\mu^+\mu^-$ (dashed lines) and
    $\tau^+\tau^-$ (dashed--dotted lines).  Constraints on
    annihilations into two quarks or two weak gauge bosons lie between
    the $e^+e^-$ and $\tau^+\tau^-$ lines.  Note that the improvements
    on the bounds between the upper and lower curves are mainly driven
    by the inclusion of $T_{\rm m}$, rather than Planck data.  A very
    similar result to the lower curves is obtained when adding the
    $T_{\rm m}$ prior to the WMAP9+SPT'11+HST+BAO data sets.}}    
\label{fig:pwave} 
\end{figure}

In Fig,~\ref{fig:pwave} we depict the $95\%$~CL limits in the
($m_\chi$, $\sigma \vref$) plane on p--wave DM annihilation cross 
sections for three different channels, which bracket the limits into
other SM annihilation channels.  To avoid a very busy plot, we only
show the limits obtained with WMAP9+SPT'11+HST+BAO (black lines and
grey regions) and those with
Planck+WMAP9--low$\ell$+ACT'10+SPT'11+HST+BAO+$T_{\rm m}$ (green lines  
and regions).  As in the case of DM decays, the bounds from WMAP9 and
Planck are very similar, and the principal source of improvement
between the upper and lower limits is the addition of the prior on
$T_{\rm m}$.  This indicates that measurements of the IGM temperature
provide a powerful tool to constrain late--time energy injection
mechanisms.  In spite of this, the annihilation cross sections probed
in Fig.~\ref{fig:pwave} are still many orders of magnitude above the
cross section required for thermal production ($b \sim 2 \times
10^{-25}$~cm$^3$/s), making these exclusions specifically relevant for
DM which was not produced via a standard chemical freeze--out.
Nevertheless, we note that in the mass range studied here, these
constraints are already stronger than the general bound from unitarity
arguments, applied to p--wave dominated annihilation cross
sections~\cite{Hui:2001wy},    
\begin{equation}
\sigma v \le 1.3 \times 10^{-12} \, \mathrm{cm}^3/\mathrm{s} \,
\left(\frac{\mathrm{GeV}}{\mx}\right)^2 \, \left(\frac{100 \mathrm{\, km
    /s}}{v_{\rm rms}}\right) ~. 
\end{equation}

Given that the dispersion velocity in halos is typically of order
$\mathcal{O}$(100~km/s), we have set $\vref = 100$~km/s, to facilitate
comparison with bounds from indirect detection.  As discussed above,
these constraints come only from the halo contribution.  Indeed, the 
$(1+z)^8/(1+z_{\rm ref})^2$ suppression to the annihilation rate of
the homogeneous cosmological DM background means that its contribution
is completely negligible as compared to the energy injected by
annihilations in halos.  This is indeed the opposite situation to what
happens in the case of s--wave annihilations.

\section{Conclusions}
\label{sec:conclusions}

Energy injection into the IGM may have important effects on the
temperature and polarization spectra of the CMB by modifying the
ionization history of the Universe after recombination.  If DM
annihilates or decays into SM particles, the produced high--energy
electrons and photons would provide an extra source of ionization and
heating that would alter the CMB.  In recent years, many analyses have
been performed to set constraints on the s--wave dominated DM
annihilation cross section using CMB data~\cite{Chen:2003gz,
  Hansen:2003yj, Pierpaoli:2003rz, Padmanabhan:2005es, Mapelli:2006ej,
  Zhang:2006fr, Ripamonti:2006gq, Zhang:2007zzh, Chuzhoy:2007fg,
  Finkbeiner:2008gw, Natarajan:2008pk, Natarajan:2009bm,
  Belikov:2009qxs, Galli:2009zc, Slatyer:2009yq, Cirelli:2009bb,
  Kanzaki:2009hf, Chluba:2009uv, Valdes:2009cq, Natarajan:2010dc,
  Hutsi:2011vx, Galli:2011rz, Finkbeiner:2011dx, Evoli:2012zz,
  Giesen:2012rp, Evoli:2012qh, Slatyer:2012yq, Frey:2013wh,
  Cline:2013fm, Weniger:2013hja, Dvorkin:2013cga,
  Lopez-Honorez:2013cua, Ade:2013zuv}.  Indeed, these limits are very 
stringent for candidates with masses below a few tens of GeV (see
Refs.~\cite{Lopez-Honorez:2013cua, Ade:2013zuv} for the latest results
using WMAP9 and Planck data).  Likewise, late--time energy injection
would raise the temperature of the IGM and this effect can be used to
further constrain the maximum allowed amount of injected energy at low
redshifts~\cite{Cirelli:2009bb, Giesen:2012rp, Lopez-Honorez:2013cua}.

In this work, we have examined two mechanisms by which DM would
release energy into the IGM with important effects at late times: DM
decays and p--wave annihilations.  In order to set constraints, we
have used the latest available CMB data: the recent Planck
data~\cite{Ade:2013zuv}, WMAP9 polarization and temperature
data~\cite{Hinshaw:2012aka} and the high--multipole
SPT~\cite{Story:2012wx, Hou:2012xq} and ACT~\cite{Sievers:2013ica} 
data.  We have also added a prior on $H_0$ from the results of the
HST~\cite{Riess:2011yx} and from BAO
observations~\cite{Anderson:2012sa, Padmanabhan:2012hf, 
  Beutler:2011hx, Blake:2011en}.  Finally, we have also added a prior 
on the matter temperature obtained from Lyman--$\alpha$ observations
at redshifts $2<z<4.5$~\cite{Schaye:1999vr}.  Indeed, the latter prior
tightens the bounds in a very significant way. 

In Fig.~\ref{fig:decay} we show our results for DM decays which,
using CMB data and adding the prior on $T_{\rm m}$, represent a lower
bound on the DM lifetime, $\tau_\chi/f_{\rm eff, dec}(\mx) \gtrsim 4
\times 10^{25}$~s.  As decays occur increasingly with time (the DM
lifetime is larger than the age of the Universe), low--multipole
temperature and polarization measurements are the most sensitive ones
when using CMB data.

In Fig.~\ref{fig:pwave} we depict the constraints we obtain for
p--wave DM annihilation cross sections.  Like for decays, the main
effects occur at low redshifts, although for a different reason.  As
DM particles begin clustering into halos, their contribution becomes
more important than the background one.  In the case of
velocity--independent annihilation cross sections, the overall
contribution from halos is smaller than that from the smooth DM
background at early times, close to recombination, so the limits are 
set by the effects caused by the latter.  However, for
velocity--dependent annihilation cross sections, there is a further 
enhancement in the contribution from halos (see Fig.~\ref{fig:rhoeff})
due to the much larger velocity of particles in halos as compared to
that of the background DM.  Hence, the contribution from halos could
even dominate the overall energy injection and it is actually the
source driving the best limits by several orders of magnitude.      

Since the most important contribution to observable effects in either
scenario comes from late--time effects, we have found that the current
determination of the IGM temperature at $z \sim 4$ solidly strengthens
the respective constraints on the decay rate and cross section.  While 
the upcoming release of Planck polarization data should improve the
CMB constraints, more accurate measurements of the IGM temperature
would significantly improve constraints  on energy injection after
recombination from DM decays and p--wave annihilations.

\section{Acknowledgments}
We would like to thank Jens Chluba for help with the \texttt{CosmoRec}
code.  LLH is supported through an ``FWO--Vlaanderen'' post--doctoral
fellowship project number 1271513.  LLH also recognizes partial 
support from the Belgian Federal Science Policy Office through the
Interuniversity Attraction Pole P7/37 and from the Strategic Research
Program ``High--Energy Physics'' of the Vrije Universiteit Brussel.
OM is supported by the Consolider Ingenio project CSD2007--00060, by
PROMETEO/2009/116, by the Spanish Grant FPA2011--29678 of the MINECO.
SPR is supported by FPA2011--23596 of the MINECO.  ACV acknowledges
support from FQRNT and European contracts FP7--PEOPLE--2011--ITN.  OM,
SPR and ACV are also supported by PITN--GA--2011--289442--INVISIBLES. 
OM gratefully acknowledges the hospitality and financial support of
the CERN Theoretical Physics Division.

\appendix

\section{Energy deposition efficiency for decaying DM}
\label{feffappendix}

Tab.~\ref{feffdectable} presents the values of $f_{\rm eff, dec}
(\mx)$ for the three decay channels and for the specific values of the
DM mass $\mx$ that were used in our MCMC analyses.

\begin{table}[t]
\begin{tabular}{ l | l l l } \hline
$\mx$(GeV) & $e$ channel \, \, \, & $\mu$ channel  \, \, \,& $\tau$
  channel  \, \, \, \\ \hline 
0.002&0.33695&--&--\\
0.003&0.53146&--&--\\
0.005&0.70445&--&--\\
0.01&0.8469&--&--\\
0.025&0.96321&--&--\\
0.05&0.95929&--&--\\
0.1&0.92764&--&--\\
0.2&0.76683&1.5997$^*$&--\\
0.3&0.63069&0.14799&--\\
0.5&0.5216&0.12527&--\\
0.75&0.45929&0.10747&--\\
1&0.43219&0.096771&--\\
5&0.3488&0.068654&0.0473$^{**}$\\
10&0.29343&0.060441&0.039151\\
20&0.22504&0.050569&0.03421\\
30&0.19927&0.044771&0.031473\\
40&0.17951&0.040639&0.029407\\
60&0.15616&0.035291&0.026516\\
80&0.14221&0.031964&0.024513\\
100&0.13295&0.029716&0.023068\\
120&0.12932&0.02817&0.022004\\
160&0.11764&0.025714&0.020352\\
200&0.11052&0.023996&0.019219\\
240&0.10463&0.022648&0.01838\\
300&0.10038&0.021215&0.017545\\
360&0.10176&0.020275&0.017039\\
400&0.09641&0.019604&0.016679\\
460&0.098201&0.019057&0.016396\\
520&0.096794&0.018515&0.016112\\
600&0.098876&0.018084&0.015889\\
640&0.097941&0.017868&0.01577\\
720&0.097191&0.017524&0.015571\\
800&0.097653&0.017301&0.015427\\
940&0.098035&0.017017&0.015224\\
1000&0.097368&0.016897&0.015133\\
 \hline
\end{tabular}
\caption{Values of the energy deposition function $f_{\rm eff, dec}
  (\mx)$, Eq.~(\ref{eq:feffdecays}) for DM decays, for each of the three
  channels considered in this work.  ($^{*}$evaluated at $\mx =
  212$\,MeV; $^{**}$evaluated at $\mx = 3.6$\,GeV).  Values represented
  by a dash (--) are below the threshold mass $2 m_{\ell}$ to produce
  the final state particle $\ell$.  Refs.~\cite{Ciafaloni:2010ti,
    Cirelli:2010xx} only provide the spectra for DM masses above
  10~GeV, so for decays into $\tau^+\tau^-$ the values of $f_{\rm eff,
    dec} (\mx)$ are obtained via extrapolation for $\mx < 10$~GeV.}
\label{feffdectable}
\end{table}

\section{Detailed calculation of the halo function $g_{\rm
    p}(z,\mx\vref)$} 
\label{haloappendix}

The energy injection via DM annihilations taking place in halos at a
given redshift $z$ is given by 
\begin{equation}
\left(\frac{\ud E}{\ud V \ud t}\right)_{\rm halo, injected}  = 
\int \ud M \frac{\ud n (M,z)}{\ud M} 
 \, \int_0^{r_{\Delta}} \ud r \, 4 \pi r^2 \, \frac{\langle \sigma v
   \rangle}{m_\chi} \, \rho_{\rm halo}^2(r) ~, 
\label{eq:haloinjectedE}
\end{equation}
where the first integral represents the sum of the contributions from
all halos and the second integral is the contribution from a single
halo and $\rho_{\rm halo}(r)$ is the density profile.  In
Eq.~(\ref{eq:haloinjectedE}), we use the physical halo mass function
${\ud n (M,z)}/{\ud M}$ which is related to the comoving one by ${\ud
  n (M,z)}/{\ud M}=(1+z)^3{\ud n_{\rm comov} (M,z)}/{\ud M}$.  We
incorporate this contribution in the total deposited energy in the IGM
of Eq.~(\ref{eq:injectedE}) in the case of p--wave annihilation making
use of
\begin{equation}
g_{\rm p}(z,\mx, \vref) = \frac{H(z)}{(1+z)^3\sum_i \int E \frac{\ud
    N}{\ud E}\ud E} \sum_i \int \ud z' \, \frac{(1+z')^2}{H(z')} \,
G_{\rm p}(z', \vref) \, \int T_i(z',z,E) \, E \, \frac{\ud N}{\ud E}
\ud E ~,
\label{eq:gofz}
\end{equation}
that depends on the dimensionless function $G_{\rm p}(z,\vref)$,
defined in Eq.~(\ref{eq:Gv}) and which we reproduce here,
\begin{equation}
G_{\rm p}(z,\vref) \equiv \frac{1}{\left(\Omega_{\rm DM,0} \,
  \rho_{c,0}\right)^2} \, \frac{1}{(1+z)^6} \, \int \ud M \, \frac{\ud
  n(M,z)}{\ud M}  \, \int_0^{r_{\Delta}} \ud r \, 4 \pi r^2 \,
\frac{\langle v^2(r) \rangle}{\vref^2}\rho_{\rm halo}^2(r) ~. 
\label{eq:Gvapp}
\end{equation}
The squared dispersion velocity of DM particles in the halo is
$\langle v^2(r) \rangle$, which depends on the DM location inside the
halo is due to the p--wave dependence of the annihilation
cross section parametrized as $\langle \sigma v \rangle= \sigma \vref
\frac{\langle v^2(r) \rangle}{\vref^2}$.  As in
Ref.~\cite{Lopez-Honorez:2013cua}, where we refer the reader for
further details, we have used the results from
Ref.~\cite{Watson:2012mt} for the halo mass function and from
Ref.~\cite{Prada:2011jf} for the halo mass--concentration relation and
have assumed an NFW DM density profile~\cite{Navarro:1995iw} for each
individual halo,
\begin{equation}
\rho_{\rm halo}(r) = \rho_s \, \frac{4}{(r/r_s) \, (1+r/r_s)^2} ~,
\label{eq:NFW}
\end{equation} 
where $r_s$ is the scale radius and $\rho_s$ the density at that
radial distance.  

In order to compute the annihilation rates in halos we assume that
halo formation is adiabatic, \textit{i.e.}, that particles remain
thermal as they contract and have a Maxwell--Boltzmann velocity
distribution    
\begin{equation}
f(v,\Sigma) = \frac{4\pi}{(2 \pi \Sigma^2)^{3/2}}v^2 \exp
\left(-\frac{1}{2}\frac{v^2}{\Sigma^2} \right), 
\end{equation}
where $\Sigma$ is the one--dimensional velocity dispersion (we use the
notation $\Sigma$ to avoid confusion with $\sigma$, the cross
section).  The squared velocity dispersion is then  
\begin{equation}
\langle v^2(r) \rangle = 3 \Sigma^2(r).
\end{equation}
If we assume hydrostatic equilibrium, the velocity dispersion can be
found by integrating the Jeans equation
\begin{equation}
\frac{\ud (\rho \Sigma^2)}{\ud r } = -\rho \frac{GM(<r)}{r^2}.
\end{equation}
This can be done analytically with an NFW profile.  The resulting
distribution of the velocity dispersion is 
\begin{eqnarray}
\Sigma^2(x) &=& \frac{8 \pi  G \rho_s r_s^2}{x} \Bigl[6 x^2 (x+1)^2
  \text{Li}_2(-x)+(x+1) \Bigl\{3 (x+1) x^2 \ln ^2(x+1) \nonumber\\ 
&+&(x+1) x^2 \Bigl(6 \ln \Bigl(\frac{1}{x}+1\Bigr)+5 \ln
  (x)\Bigr)-(x (x (5 x+11)+3)-1) \ln (x+1)\Bigr\} \nonumber \\ 
&+& x \Bigl(x \Bigl(\pi ^2 (x+1)^2-7
  x-9\Bigr)-1\Bigr)\Bigr] ~, 
\label{eq:sigsquared} 
\end{eqnarray}
where $x \equiv r/r_s$ and Li$_{2}(x)$ is the dilogarithm function,
the $n = 2$ case of the polylogarithm 
\begin{equation}
\mathrm{Li}_n(z) \equiv \frac{1}{\Gamma(n)}\int_0^\infty
\frac{t^{n-1}}{\frac{e^t}{z} -1 }\ud t ~. 
\end{equation}
We have checked that the assumption in Eq.~(\ref{eq:sigsquared}) is in
reasonable agreement with the results of the N--body simulation Via
Lactea II~\cite{Kuhlen:2009kx}.

\begin{figure}[t]
\includegraphics[width=.6\textwidth]{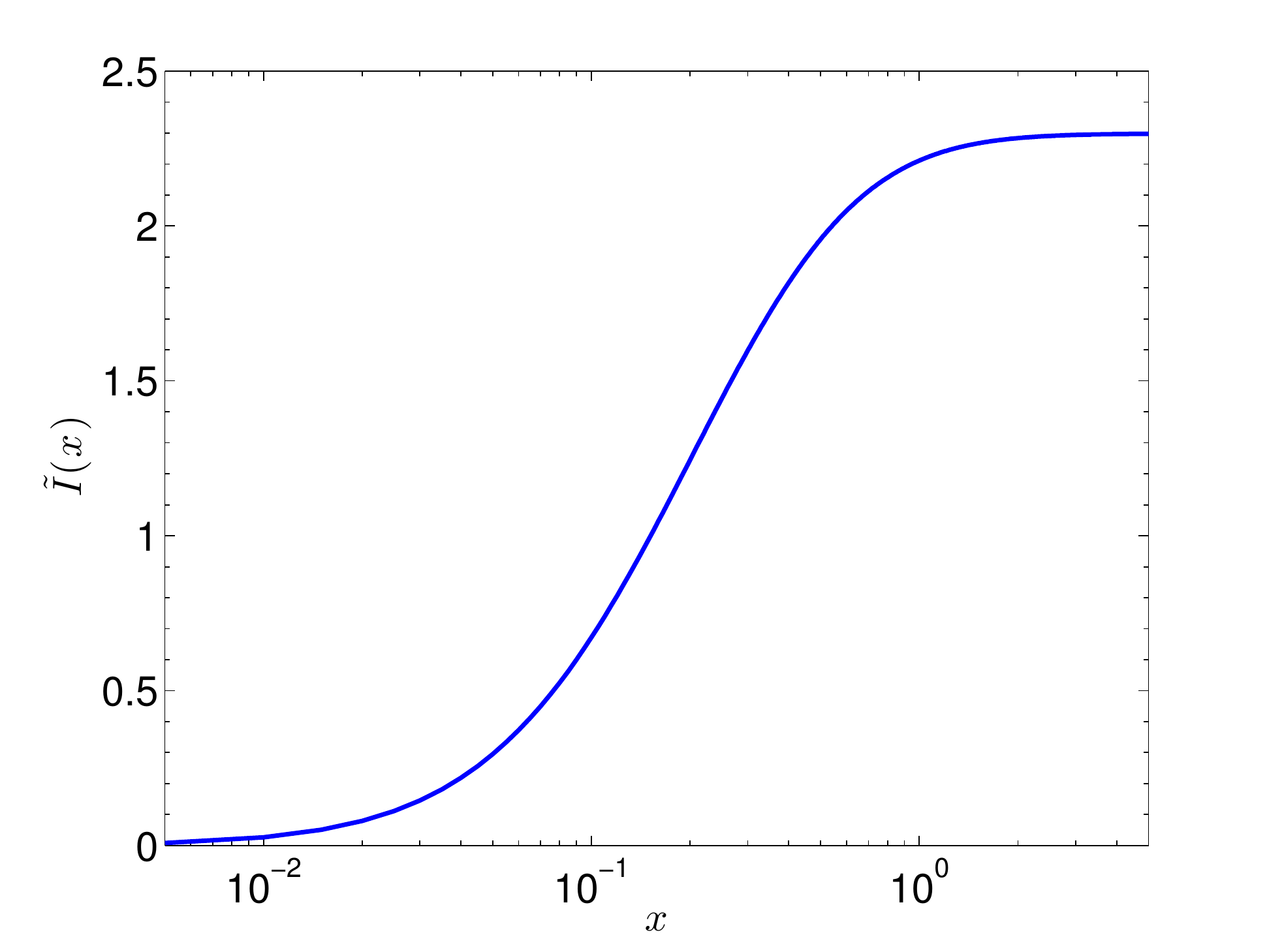}
\caption{\it The integral $\tilde I(x)$ of Eq.~(\ref{eq:bigintegral}).}
\label{fig:integralfigure}
\end{figure}

With these ingredients and the concentration parameter defined as
$c_{\Delta} = r_{\Delta}/r_s$, we can rewrite Eq.~(\ref{eq:Gvapp}) as
\begin{equation}
G_{\rm p}(z,\vref) \equiv \frac{8 \pi G r_s^5
  \rho_s^3}{\left(\Omega_{\rm DM} \, \rho_{c}\right)^2} \,
\frac{1}{(1+z)^6} \, \int \ud M \, \frac{\ud n(M,z)}{\ud M}  \,
4 \pi r^2 \tilde I(c_{\Delta}) ~, 
\label{eq:GvappI}
\end{equation}
where we have defined the dimensionless integral
\begin{equation}
\tilde I(x) \equiv \int_0^x z^2 \tilde \rho^2(z) 3 \tilde
\Sigma^2(z)\ud z = \frac{1}{8 \pi G r_s^5 \rho_s^3} \,
\int_0^{r_{\Delta}} \ud r \, r^2 \, \rho_{\rm halo}^2(r) \,
\langle v^2(r) \rangle ~,
\label{eq:I}
\end{equation}
with
\begin{equation}
\tilde \rho = \frac{\rho}{\rho_s} \hspace{5mm} ; \hspace{5mm} 
\tilde \Sigma^2 = \frac{\Sigma^2}{8 \pi G \rho_s r_s^2} ~.
\end{equation}
Thus, Eq.~(\ref{eq:I}) is explicitly given by
\begin{eqnarray}
\tilde I(x) &=& -16\Bigl[12 \text{Li}_2\left(-\frac{1}{x}\right)+36
  \text{Li}_3(x+1)-\frac{18 \text{Li}_2(x+1) (2 x+(x+1) \ln
    (x+1)+1)}{x+1} \nonumber \\ 
&+&\frac{3 x \left(6 x^2-3 x+2 \pi ^2
    (x+1)^2-21\right)}{(x+1)^3}-\frac{35}{(x+1)^3}-\frac{9 (-3 x-2)
    \ln ^2(x+1)}{x+1}\nonumber \\ 
&+&3 (\ln (x)-\ln (x+1))^3+9 \ln (x+1) (\ln (x)-\ln
  (x+1))^2+\frac{9 \ln (x+1)}{(x+1)^3} \nonumber \\ 
&-&\frac{3 \left(\pi ^2 (x+1)^3-x (2 x (5 x+11)+15)+(x+1)^3 \ln (x)+6
    (2 x+1) (x+1)^2 \ln (-x)\right) \ln (x+1)}{(x+1)^3}\nonumber
  \\ 
&-&\frac{3 (\ln (x+1)-\ln (x)) \left(\left(9+\pi ^2\right) x-4 (x+1)
    \ln (x)+6 (x+1) \ln (x+1)+\pi ^2+15\right)}{x+1} \nonumber \\ 
&-&\frac{3 \ln (x) \left(\left(10+\pi ^2\right) x+(x+1) \ln (x)
    (\ln (x)+2)+\pi ^2+15\right)}{x+1}-36 \zeta (3)+5 \pi
  ^2+35\Bigr] ~. 
\label{eq:bigintegral}
\end{eqnarray}
For $z > 0$, $\ln(-z)$ and Li$_{n}(z)$ are complex.  Taking consistent
branch cuts, and noting that
\begin{equation}
\mathrm{Im} \{\mathrm{Li}_n(z + i\epsilon)\} = \frac{\pi
  \ln^{n-1}(z)}{\Gamma(n)} ~, 
\end{equation}
one can see that the imaginary parts of Eq.~(\ref{eq:bigintegral})
cancel, leaving a real expression.  This expression is plotted in
Fig.~\ref{fig:integralfigure}.

\begin{table}[t]
\begin{tabular}{l |  l l l} \hline
$\mx$ (GeV) & $e$ channel & $\mu$ channel & $\tau$ channel \\ \hline
0.002& 12.6886 & -- & -- \\
0.005& 17.5236 & -- & -- \\
0.01 & 19.2091 & -- & -- \\
0.05& 20.3546 &-- 	& -- \\
0.105&26.3105 &7.572 	& -- \\
0.25&32.6992	 &10.073 	& -- \\ 
0.5& 22.6522	 &10.519 	& -- \\
0.75& 16.7136	 &9.3534 	& --  \\ 
1&  	13.4614	&8.2243 	& -- \\ 
1.8 & 9.2348 & 5.8579 & 5.9211 \\
5&5.7843&3.2559&3.2101 \\
25&5.9451&2.0627&1.7675\\
50&5.664&2.0536&1.4973\\
100&4.6891&1.976&1.3591\\
260&3.506&1.6286&1.2262\\
500&3.0174&1.4131&1.1676\\
1000&3.1416&1.2796&1.1168 \\ \hline
\end{tabular}
\caption{\textit{Value of $\gamma_{\rm{p} (i,\mx)}$ to be inserted into
    Eq.~(\ref{eq:fGeq}) to approximate $g_{\rm p}(z,\mx,\vref)$.
    Refs.~\cite{Ciafaloni:2010ti, Cirelli:2010xx} only provide the
    spectra for DM masses above 5~GeV, so for annihilations into
    $\tau^+\tau^-$ the value of $\gamma_{\rm{p} (i,\mx)}$ is obtained
    via extrapolation for $\mx = 1.8$~GeV.}}
\label{tab:a0table}
\end{table}

For relevant values of $z \lesssim 50$, we obtain the fit
\begin{equation}
g_{\rm p}(z,\mx,100 \mathrm{\, km /s}) \simeq f_{\rm s}(z, \mx) G_{\rm
  p}(z, 100 \mathrm{\, km /s}) \simeq \gamma_{\rm{p} (i,\mx)} \,
\Gamma_{\rm p}(z, 100 \mathrm{\, km /s}) ~,  
\label{eq:fGeq}
\end{equation}
where $\gamma_{\rm{p} (i,\mx)}$ depends on both the annihilation
channel ($i = \{e,\mu,\tau\})$ and on the DM mass $\mx$, see 
Tab.~\ref{tab:a0table}.  The function $\Gamma_{\rm p}(z, 100
\mathrm{\, km /s})$ is given by  
\begin{equation}
\ln{\Gamma_{\rm p}(z, 100 \mathrm{\, km /s})} = a_3 (1+z)^3 + a_2
  (1+z)^2 + a_1 (1+z) + a_0 
\end{equation}
with 
\begin{equation}
a_3 = -7.379 \times 10^{-5} \; ; \; \; \; 
a_2 = - 0.004499 \; ; \; \; \; 
a_1 = -0.7012 \; ; \; \; \; 
a_0 = 4.2 ~. 
\end{equation}

Finally, we have 
\begin{equation}
g_{\rm p}(z,\mx,\vref) = \left(\frac{100 \mathrm{\,
    km/s}}{\vref}\right)^2 \, g_{\rm p}(z,\mx,100 \mathrm{\, km/s}) ~. 
\label{eq:gp}
\end{equation}

\bibliographystyle{utphys}
\bibliography{efficiency.bib}

\end{document}